\documentclass[12pt]{article}
\usepackage{graphicx}
\usepackage{amssymb,latexsym,amsmath}
\begin{document}
\title{ Counterion atmosphere around DNA double helix: trapping of counterions at the nanoscale}
\author{Sergiy Perepelytsya\\
Bogolyubov Institute for Theoretical Physics, NAS of
Ukraine,\\14-b Metrolohichna Str., Kiev 03143, Ukraine \\
perepelytsya@bitp.kiev.ua \\
Oleksii Zdorevskyi\\
Department of Physics, University of Helsinki, FI-00014 Helsinki, Finland}\maketitle

\setcounter{page}{1}%
\maketitle
\abstract{DNA is strong polyelectrolyte macromolecule making metal ions (counterions) condense to a cloud around the double helix.  The counterions may be localized outside the macromolecule and inside the minor and major grooves of the double helix. In the present work, the distribution of condensed counterions between inner and outer regions of DNA has been studied using the approaches of counterion condensation theory. The results have shown that the number of counterions trapped inside the macromolecule should be greater than 0.16 per one phosphate group. The maximal number of counterions that may be localized inside the DNA double helix is limited to about 0.4 per one phosphate group and it is much lower than the total number of condensed counterions. To analyze the structure of counterion cloud the molecular dynamics simulations of \emph{B}-DNA with K$^{+}$ counterions have been performed. The obtained number of the counterions trapped inside the grooves of the double helix is about 0.22$\pm$0.06 per one phosphate group that agree with the model estimations. The developed model describes general features of the structure of counterion cloud around DNA and is able to predict the number of counterions inside the grooves of the double helix.
}

\maketitle

\textbf{Keywords:} DNA, counterion condensation, polyelectrolyte, molecular dynamics.

%
\section{Introduction}
\label{intro}
DNA is the charged macromolecule consisted of two strands of twisted nucleotide chains in a shape of the double helix. The nucleotides form the H-bonded complementary base pairs inside the double helix and regular sugar-phosphate backbone outside the macromolecule \cite{Saenger}. In aqueous solutions each phosphate group of the macromolecule backbone has one excess electron, which gives the DNA polyanion properties. The negative charge of DNA is neutralized by the positively charged ions of the solution (counterions) stabilizing the structure of the double helix. Under the physiological conditions, the metal ions, like Na$^+$, K$^+$, Mg$^{2+}$, and organic molecules, like spermidine$^{3+}$ and spermine$^{4+}$ play the role of neutralizing counterions \cite{Blagoy}. The counterions around the macromolecule determine the conformational transformations of the double helix that are important in the mechanisms of DNA biological functioning \cite{vologodskii_biophysics_2015}.

The highly charged polyelectrolyte molecules make the counterions condense to a cloud around the macromolecule (see the reviews \cite{Frank_Kamenetski__1987,levin_electrostatic_2002,kornyshev_structure_2007}). The effect of counterion condensation for different polyelectrolytes was predicted within the simples models of counterion condensation theory \cite{manning_limiting_1969,oosawa_counterion_1970,osawa_polyelectrolytes_1971,manning_1978}. In these models the DNA macromolecule is presented as a chain of charged beads surrounded by counterions reducing the charge of the macromolecule \cite{manning_1978}. For the case of the natural \emph{B}-form of the double helix, the number of counterions per one phosphate group in the counterion cloud is about 0.76, and the thickness of counterion cloud is about 7 {\AA} \cite{manning_1978}. The manifestations of the effects of counterion condensation and ordering around the DNA double helix have been observed in small angle X-ray scattering  \cite{Das,Andresen}, and other experiments, for example \cite{tomic_dielectric_2007,perepelytsya_texture_2013,Liubysh2014}. The basic polyelectrolyte models have been modified for the description of the double helix bending, persistence length, polarizability of  counterion cloud and others effects \cite{manning_persistence_2006,manning_contribution_2006}.

In the following polyelectrolyte models the DNA macromolecule is usually presented as a uniformly charged cylinder surrounded by the positively charged continuum of the ions \cite{Frank_Kamenetski__1987}. The  models are analyzed within the framework of Poisson-Boltzmann (PB) equation. As the result, the distribution of counterions as the function of distance to the DNA surface was calculated, and in some cases even the analytical solutions were obtained \cite{Katchalsky,Obukhov,oshaughnessy_manning-oosawa_2005,trizac_onsager-manning-oosawa_2006}. The models based on PB equation give qualitatively the same conclusion related to the counterion condensation, but the concentration of the counterions in a cloud around DNA gradually decreases to the bulk concentration as the distance from macromolecule increases \cite{Frank_Kamenetski__1987}. In this case  there is no defined border between counterion cloud and solution that is obtained in simple models of counterion condensation theory \cite{manning_1978}. In spite of the progress of polyelectrolyte theories of DNA, the structure of counterion cloud is still far from complete understanding.

The counterion cloud may be conditionally divided into the internal and external regions of the DNA macromolecule. The minor groove and major groove of the double helix may be considered as the inner regions of DNA, while the shell of counterion cloud between DNA surface and solution may be considered as the outer region. According to molecular dynamics simulations, the  residence time of counterions inside DNA is within the range from 10 ps to 1 ns, depending on counterion type, compartment of the double helix, sequence of nucleotide bases, and character of hydration \cite{Mocci,mocci2012,Aksimentiev2012,Lavery2014,Canadian,dans2016,Maddocks,Perepelytsya2018,perepelytsya2019positively}. Within these residence times, the counterions can do the vibrations that are observed in the low-frequency vibration spectra of different forms of the DNA double helix; in particular, the vibrations of counterions relatively the phosphate groups (ion-phosphate vibrations) have been determined in the  spectra range from 90 to 180 cm$^{-1}$  \cite{PV_EPJE_2007,PV_EPJE_2010,PV_JML_2011,Bulavin,perepelytsya_BZ_2013,perepelytsya_left_2013}. The counterions outside the double helix move around the macromolecule without constraints. Thus, the structure of counterion cloud around DNA is not uniform and has different properties inside and outside the double helix. To describe the distribution of counterions by the compartments of double helix new models should be elaborated.

The goal of the present work is to study the structure of the counterion cloud around DNA and to characterize the fraction of condensed counterions inside and outside the double helix. The structure of the present work is as follows. In the Section 2, the basics of simples model of counterion condensation theory \cite{manning_1978} have been described. In the Section 3, a polyelectrolyte model has been developed on the basis of the approach \cite{manning_1978} for the case of monovalent counterions. Using the developed model the estimations of the number of condensed counterions inside and outside the double helix have been performed for the case of \emph{B}-forms of DNA double helix. In the Section 4, the molecular dynamics simulations have been carried out to analyse the distribution around the DNA double helix, and the comparison of the model and simulation results have been performed.

\section{Basics of counterion condensation theory}
In the present work the initial model of counterion condensation theory \cite{manning_1978} is extended to take into consideration the effect of trapping of counterions inside DNA. Further, the basics of counterion condensation theory \cite{manning_1978} will be described in brief.

Taking into consideration the structure of the DNA double helix (Fig. 1a), the initial model of the counterion condensation theory presents the DNA macromolecule as a chain of charged beads separated by the distance $b$ (Fig. 1b). Each charge of the bead equals to the value of elementary charge ($q$). The distance $b$ between charges is such as to reproduce the linear charge density of the macromolecule. In the case of the DNA double helix, the distance between charges corresponds to the half of the distance between nucleotide pairs, since one nucleotide pair contains two phosphate groups, having the negative charge equal to $q$. The counterions around the DNA macromolecule are  modeled as a charged continuum. The ions that are localized in some volume $V_p$ around the charged chain are considered as condensed counterions. Condensed counterions screen the electrostatic charge of the macromolecule. The fraction of condensed counterions per one phosphate group is described by the parameter $\theta$. Each charge of the polyelectrolyte chain is reduced to the value: $q(1-\theta)$.

\begin{figure}
\begin{center}
\resizebox{0.5\columnwidth}{!}{
\includegraphics{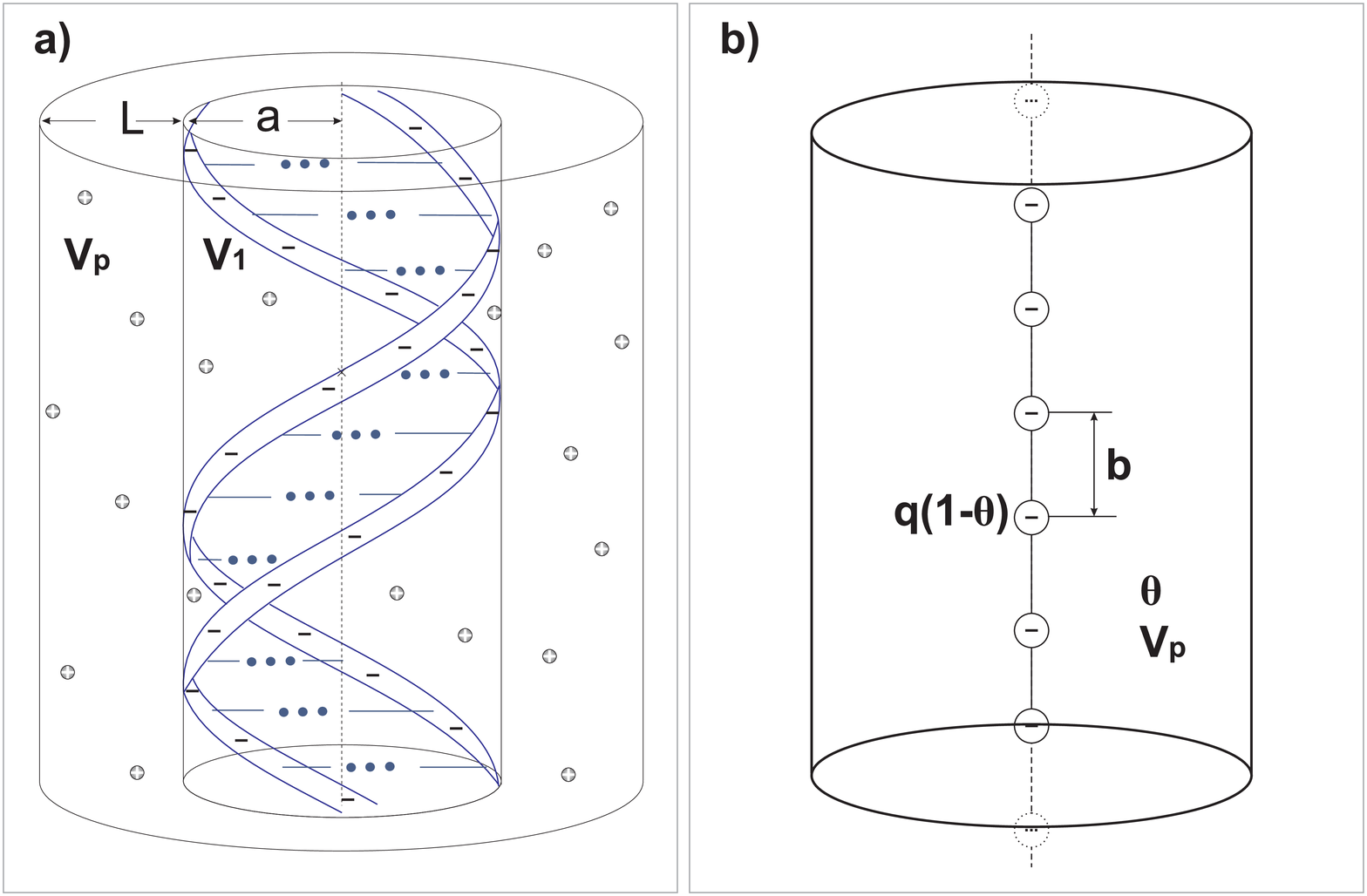}
}
\end{center}
\caption{a) The DNA double helix with counterions neutralizing the negatively charged phosphate groups. b) Initial model of counterion condensation theory \cite{manning_1978}: $q$ is the elementary charge, $b$ is the distance between charges that corresponds to the linear charge density of the polyanion; $\theta$ is the number of condensed counterions per one phosphate group; $V_p$ is the volume per one phosphate group accessible for the counterions.}
\end{figure}

Free energy of the system is presented as a sum of electrostatic $g_{rep}$ and entropy $g_{mix}$ contributions \cite{manning_1978}:
\begin{equation}\label{Eq1}
g=g_{rep}+g_{mix}.
\end{equation}
Here, the free energy is in units of $k_BT$ per one phosphate group, $k_B$ is the  Boltzmann constant, $T$ is the temperature in Kelvins.

In the counterion condensation theory \cite{manning_1978}, the mean electrostatic energy of repulsion between the charged beads was analyzed using the Coulomb potential with the the Debye screening. The charge of each phosphate group was reduced by the number of condensed counterions $\theta$ (Fig. 1b). As the result, the electrostatic energy was obtained in the following form:
\begin{equation}\label{Eq2}
 g_{rep}=-\xi(1-\theta)^2\ln(\kappa{b}).
\end{equation}
Here $\xi$ is the dimensionless linear density of charge of the polyion that may be written as $\xi=L_B/b$. The parameter $L_B=q^2/4\pi\varepsilon\varepsilon_0k_BT$ is the Bjerrum length, $\varepsilon$ is the dielectric constant of the solution, $\varepsilon_0$ is the dielectric constant of vacuum. The Bjerrum length is the distance at which the energy of interaction of two electrostatic elementary charges equals to the thermal energy. The parameter $\kappa$ is the transverse Debye length: $\kappa=\sqrt{8L_B\times10^3{N_A}^2C}=A\sqrt{C}$, where $N_A$ is the Avogadro constant, and  $C$ is the concentration of the ions in moles per liter.

The term $g_{mix}$ in (1) describes the entropy of mixing of counterions in a cloud around the macromolecule with the counterions in the solution. It is determined by the concentration of condensed counterions and by the concentration of the ions in solution. Taking into consideration that the number of counterions per each chain bead is $\theta$, the term of mixing entropy may be written in the following form \cite{manning_1978}:
\begin{equation}\label{Eq3}
 g_{mix}=\theta\ln{\frac{\theta}{V_{p}C}}.
\end{equation}
In the logarithm the ratio of the concentrations is present: $\theta/V_p$ is the concentration of condensed counterions and $C$ is the concentration of salt in the solution. Since the concentration of the ions $C$ is usually in moles per liter, the volume $V_p$  is taken in liter per mole.

Minimizing the free energy ($dg/d\theta=0$) the equation with respect to $\theta$ was obtained. To  omit the singularity that appeared due to the $\ln{C}$ in the case of $C\rightarrow0$, the coefficient near the $\ln{C}$ was equaled to zero. As the result the equilibrium value of the  parameter $\theta$ is obtained in the following form \cite{manning_1978}:
\begin{equation}\label{Eq4}
\theta=1-\xi^{-1}.
\end{equation}

Using the condition of energy minima and the equation (4) the formula for the volume was obtained in the following form \cite{manning_1978}:
\begin{equation}\label{Eq5}
V_p=A^2b^2e(1-\xi^{-1}).
  \end{equation}
The counterions are considered condensed in the case if they are localized in a volume $V_p$ around the phosphate group.
The estimations show that the thickness of the counterion shell should be about 7 {\AA} that is close the the value of Bjerrum length.

As follows from (4), the cloud of counterions around the polyelectrolyte molecule is formed in the case $\xi>1$ \cite{manning_1978}. This condition means that to observe the condensation of countreions the distance between elementary charges in the model must be lower than the Bjerrum length ($b<L_B$). Taking into consideration the value of Bijerrum length for water solution (about 7.4 {\AA} at 300 K) and the structural parameters of the DNA macromolecule, the effect of counterions condensation is expected to be observed for the case of any form of the DNA double helix. It should be noted that in the basic model of counterion condensation theory the counterion cloud is uniform and the inner and outer regions of the DNA double helix do not differ.

\section{Trapping model of condensed counterions}

The structure of the DNA double helix is characterized by the minor groove and major groove, where the counterions may be localized (Fig. 1a). To take into consideration trapping of counterions inside the DNA double helix, the initial model of counterion condensation theory \cite{manning_1978} described above  has been extended in the present work.

In the developed model, DNA macromolecule is presented as the infinite chain of charged beads separated by the distance $b'$. The total number of condensed counterions per one phosphate group is $\theta$. The counterions are considered to be condensed on DNA if they are localized inside the volume $V_p$ around the polyelectrolyte. The inner volume of the DNA double helix is presented by the volume $V_1$ around the polyanion. The counterions inside this volume (trapped counterions) reduce the linear charge density of the macromolecule. The ability of polyanion to trap the counterions inside is described by the trapping parameter $\gamma$ that is the number of trapped counterions per one phosphate group (Fig. 2).
\begin{figure}
\begin{center}
\resizebox{0.3\columnwidth}{!}{
\includegraphics{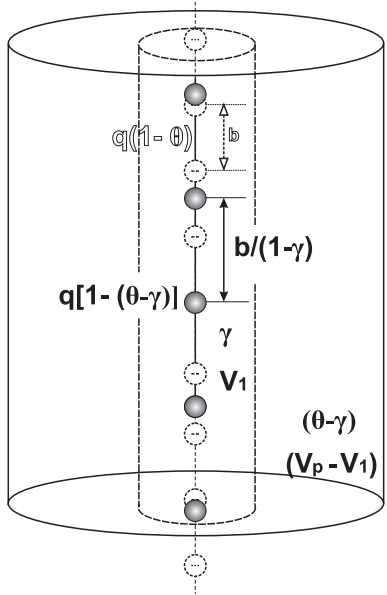}
}
\end{center}
\caption{Trapping model of condensed counterions. $b'$ is the distance between renormalized charges; $\theta$ is the number of condensed counterions per one renormalized charge; $V_1$ is the volume per one phosphate group accessible for counterions inside DNA; $(V_p-V_1)$ is the volume per one renormalized charge accessible for the counterions outside the macromolecule; $\gamma$ is the trapping parameter, $\gamma$ and $(\theta-\gamma)$ are the number of counterions per one phosphate group inside and outside the macromolecule, respectively. The position of charges at the initial separation distances $b$ are shown as dashed circles. }
\end{figure}

Due to the neutralization of the phosphate groups by the counterions that are localized inside the macromolecule, the linear charge density is reduced. Thus, the number of charged beads in our model $N'$ decreases by the number of trapped counterions: $N'=N-N\gamma$, where $N$ is the number of phosphate groups in DNA molecule (in our model $N\rightarrow\infty$). Taking into consideration that $b'=L/(N'-1)$ and $b=L/(N-1)$, the  distance between charges in our model may be found in the following form:
\begin{equation}\label{Eq6}
  b'=b/(1-\gamma).
\end{equation}

The charged beads separated by the distance $b'$ are screened by the counterions that are localized outside the double helix in the counterion cloud. The number of counterions outside the double helix per one phosphate group is $(\theta - \gamma)$. Thus, taking into consideration the formula (6), the mean electrostatic energy of repulsion of the charged beads between each other (2) may be rewritten in the following form:
\begin{equation}\label{Eq7}
 g_{rep}=-\xi(1-\gamma)[1-(\theta-\gamma)]^2\ln\left(\frac{\kappa{b}}{1-\gamma}\right).
\end{equation}

The entropy mixing contribution has been determined, considering the counterions localized in the regions outside the macromolecule in the volume $(V_p-V_1)$ and inside macromolecule in the volume $V_1$ (Fig. 2b). The volume of the inner part of the DNA double helix $V_1$, where the trapped counterions are localized, is the characteristic of the structure of the double helix. Thus, the part of free energy, related to the mixing entropy (3), may be rewritten in the following form:
\begin{equation}\label{Eq8}
 g_{mix}=\theta\ln\frac{\theta}{V_{p}C}+(\theta-\gamma)\ln\frac{(\theta-\gamma)V_p}{(V_{p}-V_1)\theta} +\gamma\ln\frac{\gamma{V_p}}{V_1\theta}.
\end{equation}
As before $C$ is the molar concentration in mole per liter, and $V_p$  is the volume that occupy the condensed counterions per one phosphate group in liter per mole. The first term in the equation (8) describes the entropy due to the mixing of all condensed counterions with the ions of the solution, and it is the same as in the case of Manning model [equation (3)]. The second and the third terms in (8) are the entropy of mixing of the ions of whole cloud with the counterions that are localized in the cloud outside and inside DNA macromolecule, respectively.

Taking into consideration the electrostatic (7) and entropy (8) contributions, the free energy of the system of counterions around the polyelectrolyte molecule is determined as a function of parameter $\theta$. The equilibrium value of parameter $\theta$ has been determined from the condition of energy minima $dg{/}d\theta=0$ that may be written in the following form:
\begin{equation}\label{Eq9}
G(\theta)+F(\theta)\ln{C}=0.
\end{equation}
Here $G(\theta)$ and $F(\theta)$ denote the following functions:
\begin{equation}\label{Eq10}
\begin{split}
G(\theta)=\xi[2(1-\gamma)-\gamma'(1-3\gamma+\theta)][1-(\theta-\gamma)]\ln\frac{A{b}}{1-\gamma}+\\
\ln{\frac{(\theta-\gamma)e}{V_p-V_1}}-\xi\gamma'[1-(\theta-\gamma)]^2+\gamma'\ln{\frac{\gamma(V_p-V_1)}{(\theta-\gamma)V_1}},\\
\end{split}
\end{equation}

\begin{equation}\label{Eq11}
\begin{split}
F(\theta)=\frac{\xi}{2}[1-(\theta-\gamma)][2(1-\gamma)-\gamma'(1-3\gamma+\theta)]-1,
\end{split}
  \end{equation}
where $\gamma'=d\gamma/d\theta$, $e$ is the base of natural logarithm.

To remove the singularity in the case of zero concentration of salt ($C\rightarrow0$) the coefficients near the $\ln{C}$ in the equations (9) should be put to zero, the same as in counterion condensation theory \cite{manning_1978}. Taking this into consideration, the functions $G(\theta)$ (10) and $F(\theta)$ (11) should be put to zero, since they are the parts of the same equation (9). As the result, the following equations may be obtained:
\begin{equation}\label{Eq12}
\begin{split}
2\ln\frac{A{b}}{1-\gamma}+\ln{\frac{(\theta-\gamma)e}{V_p-V_1}}-\xi\gamma'[1-(\theta-\gamma)]^2+\gamma'\ln{\frac{\gamma(V_p-V_1)}{(\theta-\gamma)V_1}}=0,\\
\end{split}
\end{equation}
\begin{equation}\label{Eq13}
\begin{split}
\gamma'\theta^2-2\theta(1-\gamma)(1+\gamma')+(1+\gamma)[2(1-\gamma)-\gamma'(1-3\gamma)]-2\xi^{-1}=0.
\end{split}
\end{equation}
Using the equations (12) and (13) the equilibrium values for the volume $V_p$ and the parameter $\theta$ may be calculated. In the case of $\gamma=0$ and $V_1=0$, the equations (12) and (13) provide the same result as in the case of the counterion condensation theory \cite{manning_1978} [see the equation (4) and (5)].

The number of counterions inside  DNA is determined by trapping parameter $\gamma$ that determines the character of counterion redistribution among the internal and external compartments of the double helix. In general, the trapping parameter $\gamma$ is the function of the parameter $\theta$. The form of this function is not defined and the analysis should be carried out to elucidate the character of its behaviour.

The minimal value of trapping function $\gamma(\theta)$ that may be considered in the model is $\gamma=0$. The case of a zero value of this function means that there is no trapping of counterions inside the polyelectrolyte. It may be as a result of absence of counterions around DNA ($\theta=0$) or other reasons related to the features of the interaction of counterions with the macromolecule. Increasing the number of counterions around the polyelectrolyte, the internal compartments of the double helix should be gradually filled with the counterions, and $\gamma$  should increase as $\theta$ increases. At the same time, the number of counterions that may be localized inside the double helix is constrained by the volume of the internal compartments ($V_1$) and by the electrostatic repulsion of the counterions between each other. Thus, the trapping parameter should reach the plateau  ($\gamma_{max}$) under the maximal concentration of condensed counterions ($\theta=\theta_{max}$). To determine $\gamma_{max}$ let us analyze the equation (13) for the case of maximal condensation of counterions that corresponds to the condition $\theta_{max}=1$. For the plateau value of trapping parameter the derivative $\gamma'=0$. Taking this into consideration the following equation with respect to $\gamma_{max}$ may be obtained from the equation (13):
\begin{equation}\label{Eq14}
\gamma_{max}^2-\gamma_{max}+\xi^{-1}=0.
  \end{equation}
The solutions of the quadratic equation (14) are as follows:
\begin{equation}\label{Eq15}
\gamma_{max}^{\pm}=\frac{1}{2}\pm\frac{1}{2}\sqrt{1-4\xi^{-1}}.
  \end{equation}
It should be noted that $\gamma_{max}^{+}>\gamma_{max}^{-}$. The solution with lower value of trapping parameter ($\gamma_{max}^{-}$) is the most interesting in our study, since in the case of $\gamma_{max}^{+}$ the polyanion is almost neutralized by trapped counterions and the effect of counterion condensation will be less prominent. To elucidate the character of counterion redistribution among the internal and external regions of the DNA double helix some particular cases of the trapping function are analyzed further.

In a simplest case, the trapping parameter may be considered as a constant that does not depend on parameter $\theta$, $\gamma(\theta)=\gamma_0$. Such approximation may be valid if the number of counterions inside the polyelectrolyte weakly depends on concentration of condensed counterions around the macromolecule. For DNA polyelectrolyte the concentration of counterions in a cloud should be much larger than in the internal compartments of the double helix, and the imposed condition may be fulfilled. Taking this into consideration, and using the equations (12) and (13), the parameter $\theta$ and the volume $V_p$ may be determined in the following form:
\begin{equation}\label{Eq16}
\theta=1+\gamma_0-\frac{1}{\xi(1-\gamma_0)},
  \end{equation}

  \begin{equation}\label{Eq17}
V_p=V_1+\frac{A^2b^2e}{(1-\gamma_0)^2}\left[1-\frac{1}{\xi(1-\gamma_0)}\right].
  \end{equation}
In this case, the both $\theta$ and $V_p$ are the functions of trapping parameter $\gamma_0$, which is unknown.

On the other hand, the assumption of a uniform distribution of counterions around the DNA double helix can be used. 
In this case, the concentration of counterions inside the double helix (in the volume $V_1$) should be equal to the average concentration of counterions in a cloud around DNA (in the volume $V_p$). To fulfill  this condition, the trapping function $\gamma=\theta{V_1}/V_p$ may be considered. As the result, the equations (12) and (13) take the following form:
\begin{equation}\label{Eq18}
2\ln\frac{A{b}}{\left(1-\theta\frac{V_1}{V_p}\right)}-\xi\frac{V_1}{V_p}\left[1-\theta\left(1-\frac{V_1}{V_p}\right)\right]^2=0,
\end{equation}
\begin{equation}\label{Eq19}
\begin{split}
3\theta^2\frac{V_1}{V_p}\left(1-\frac{V_1}{V_p}\right)^2-2\theta\left(1-\frac{V_1^2}{V_p^2}\right)+2(1-\xi^{-1})-\frac{V_1}{V_p}=0.
\end{split}
\end{equation}
The equation (19) has the following solutions:
\begin{equation}\label{Eq20}
\theta^{\pm}=\frac{1+\frac{V_1}{V_p}\pm\sqrt{\left(1+\frac{V_1}{V_p}\right)^2-3\frac{V_1}{V_p}\left(2-\frac{2}{\xi}-\frac{V_1}{V_p}\right)}}{3\frac{V_1}{V_p}\left(1-\frac{V_1}{V_p}\right)}.
  \end{equation}
Substituting the solutions (20) to the equation (18) the transcendent equation with respect to $V_p$ is obtained. The uniform distribution of counterions will be referred to the case corresponding to the minimal number of trapped counterions, which will be denoted as $\gamma_{min}$. The case when the number of trapped counterions is less than $\gamma_{min}$ corresponds to the fact that counterions must be pushed out from inside the DNA, which is not observed experimentally.
In contrast, the case when the number of trapped counterions is greater than $\gamma_{min}$ is possible when counterions get stuck in the grooves of the double helix, which is observed in molecular dynamics simulations for various counterions.  \cite{Mocci,mocci2012,Aksimentiev2012,Lavery2014,dans2016,Maddocks,Perepelytsya2018}.

To perform the estimations within the framework of the developed model the parameters of a polyanion should be determined. The parameter $b$ that determines the distance between elementary charges in the model is the half of a distance between DNA base pairs along the helical axis ($h$), and in the case of \emph{B}-DNA $b=1.7$ {\AA}. The radius of the DNA double helix was taken equal to 10.5 {\AA} that corresponds to \emph{B}-form. The parameter $V_1$ has been estimated as a half of the volume that is the difference of cylindrical disc volume around the nucleotide pair and the proper volume of the atoms of DNA nucleotides ($V_{bp}$):
\begin{equation}\label{Eq21}
V_1=(\pi{a^2}h-V_{bp})/2,
\end{equation}
where $a$ is the radius of the double helix. In the equation (21), the volume is divided by 2 because the nucleotide pair has two phosphate groups, while the $V_1$ is the volume per one phosphate group. The values for volumes of atoms in DNA nucleotide pairs were determined in \cite{nadassy_standard_2001} for different nucleotides: adenine (A), guanine (G), thymine (T) and cytosine (C). Using the data \cite{nadassy_standard_2001}, the value of the volume $V_{bp}$, required for the calculation of $V_1$ by the formula (21), has been estimated  as the average volume of A-T and G-C complementary nucleotide pairs: $V_{bp}=613$ {\AA}$^{3}$. As the result the volume  $V_1$ estimated ny the formula (21) is $V_1=283$. The dielectric constants $\varepsilon=80$ was used in calculations. The Bjerrum length for such dielectric constant is 7.0 {\AA}. Using the determined parameters and the formula (15) for $\gamma_{max}$ and formula (20) for $\gamma_{min}$ the range of variation of the trapping constant has been estimated (Table 1).

\begin{table}
\caption{The parameters of counterion cloud for \emph{B}-DNA. }
\label{tab:model}       
\begin{center}
\begin{tabular}{lcccc}
\noalign{\smallskip}\hline\noalign{\smallskip}
	Parameter	            & $\theta$   &	$\gamma$     &	$V_{p}$, l/mol  & $L$, {\AA}	\\[2mm]
\noalign{\smallskip}\hline\noalign{\smallskip}
Minimal value                &0.82       & 0.16          &0.87            & 7.6 \\
Maximal value                &1          & 0.43          &1.63	            & 13.2 \\
Manning theory    \cite{manning_1978}            &0.76       & --          &0.63            &5.4\\[2mm]
\noalign{\smallskip}\hline
\end{tabular}
\end{center}
\end{table}

The results have shown that the minimal $\theta$  value is larger than in the case of Manning theory. At the same time, the volume of counterion cloud around DNA double helix is also larger: the thickness of the counterion shell around the double helix is more than 1 {\AA} larger. At least about 20 {\%} of all counterions are trapped inside the grooves of the double helix. The estimated maximal number of trapped counterions is essentially higher, but this is the limiting case for $\theta=1$, taken to estimate the range of $\gamma$ that is from 20 to 40 {\%}.

Thus, the developed model allows to analyze trapping of counterions inside DNA double helix. The trapping parameter $\gamma$ is the key feature of the model varying within the range from $\gamma_{min}$ to $\gamma_{max}$. The upper limit of the trapping parameter $\gamma_{max}$ has been determined analytically (15) from the condition of electroneutrality. The minimal number of trapping parameter $\gamma_{min}$ has been determined from the condition of uniform distribution of condensed counterions. The distribution of the counterions between the internal and external regions of the double helix in the case of some intermediate values of $\gamma$ may be elucidated using the equations (16)  and (17).

\section{Molecular dynamics analysis}

To validate the introduced model, the atomistic molecular dynamics (MD) simulations of an infinite macromolecule with repeating 22-base-pair \emph{B}-DNA fragment d(CGCGAATTCGCGCGAATTCGCG) were carried out. The fragment of DNA double helix was immersed into a water box with the size 62$\times$62$\times$78 {\AA}$^3$. The number of water molecules was 8374. DNA with this nucleotide sequence consists of two fragments, known as Drew-Dickerson dodecamer \cite{1bna}, and is commonly used in the molecular dynamics simulations of DNA systems \cite{Charge2000,rueda2004,Ponomarev2004,mocci2012,SJAF}. The system was neutralized by KCl salt 0.15 M concentration. The initial positions of the counterions were randomly generated to be more than 5 {\AA} apart from each other and at least 7 {\AA} from the DNA double helix. In the simulations, the periodic boundary conditions were used. To produce a model of an infinite DNA duplex, the ends of the polynucleotide were linked with their images in the adjacent boxes.

\begin{figure}
\begin{center}
\resizebox{0.5\columnwidth}{!}{
\includegraphics{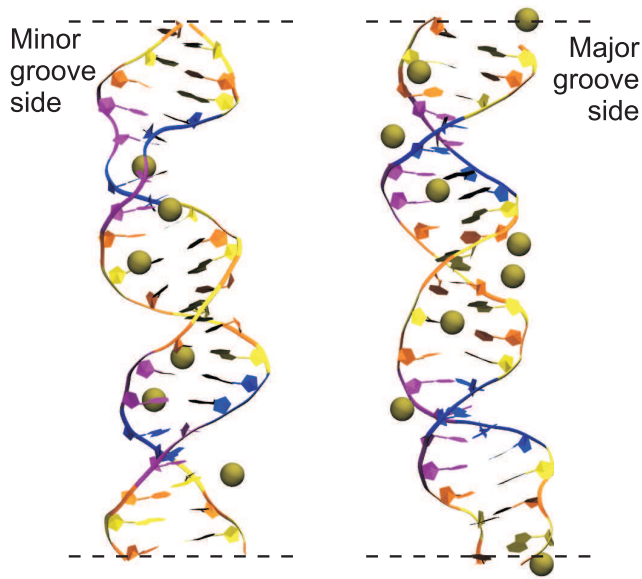}
}
\end{center}
\caption{DNA with counterions in the minor and major grooves of the double helix. The dotted lines indicate the boundaries with the image box. Coloring scheme corresponds to the type of nucleotides: Cytosine (orange), Guanine (yellow), Adenine (blue), Thymine (purple).}
\end{figure}

The computer simulations were performed using NAMD software package \cite{Phillips} and CHARMM27 force field \cite{MacKerell2000,Foloppe2000}. The VMD program was used for the systems construction, visualization and analysis~\cite{Humphrey}. All covalent bonds to hydrogen atoms were constrained via SHAKE algorithm \cite{SHAKE}. To link  the ends of the DNA fragment with their images in the adjacent boxes the patch LKNA from CHARMM33 parameter set was taken. The TIP3P water model \cite{TIP3P} and the Beglov and Roux parameters of ions  \cite{Beglov} were used. The Langevin dynamics was applied to all heavy atoms with the temperature 300 K. The long-range electrostatic interactions were treated by particle mesh Ewald method \cite{PME}. The switching and cutoff distances for the long-range interactions were 8 {\AA} and 10 {\AA}, respectively.

The simulation protocol was taken the same as in previous molecular dynamics studies of DNA with counterions \cite{Zdorevskyi2020}. K-DNA system was initially simulated at the constant pressure 101325 Pa and temperature 300 $^{\circ}$K (NPT ensemble). The procedures of minimization, heating and equilibration were performed for the system with restrained DNA atoms. The restraints were made gradually weaker, and after about 5 ns of equilibration all DNA atoms were free to move. Then the system was simulated in NVT ensemble for 300 ns. For the counterion analysis, the last 100 ns of the production run were taken.

Visual inspection of the simulation trajectories shows that  K$^{+}$ counterions interact with DNA at different binding sites: minor and major grooves, as well as phosphate groups regions. The most important feature is that K$^{+}$ penetrate deep inside the minor groove, interacting with the atoms of nucleotide bases, and stay there for a rather long time (up to nanoseconds). The snapshots from the simulation trajectory for DNA with counterions in the minor and major grooves of the double helix at 220 ns and 212 ns, respectively, are shown in Figure 3.

The effect of counterion trapping in the minor groove was studied in \cite{Perepelytsya2018,perepelytsya2019positively}, where the character of ion hydration was found crucial in this case. The potassium ions may be easily dehydrated, therefore they squiz through the hydration shell of the macromolecule to the bottom of the groove \cite{Perepelytsya2018}. The position of potassium ions in the minor groove is so favorable that counterions follow along the minor groove even after applying the electric field that was studied in \cite{Zdorevskyi2020}.

\begin{figure}
\begin{center}
\resizebox{0.5\columnwidth}{!}{
\includegraphics{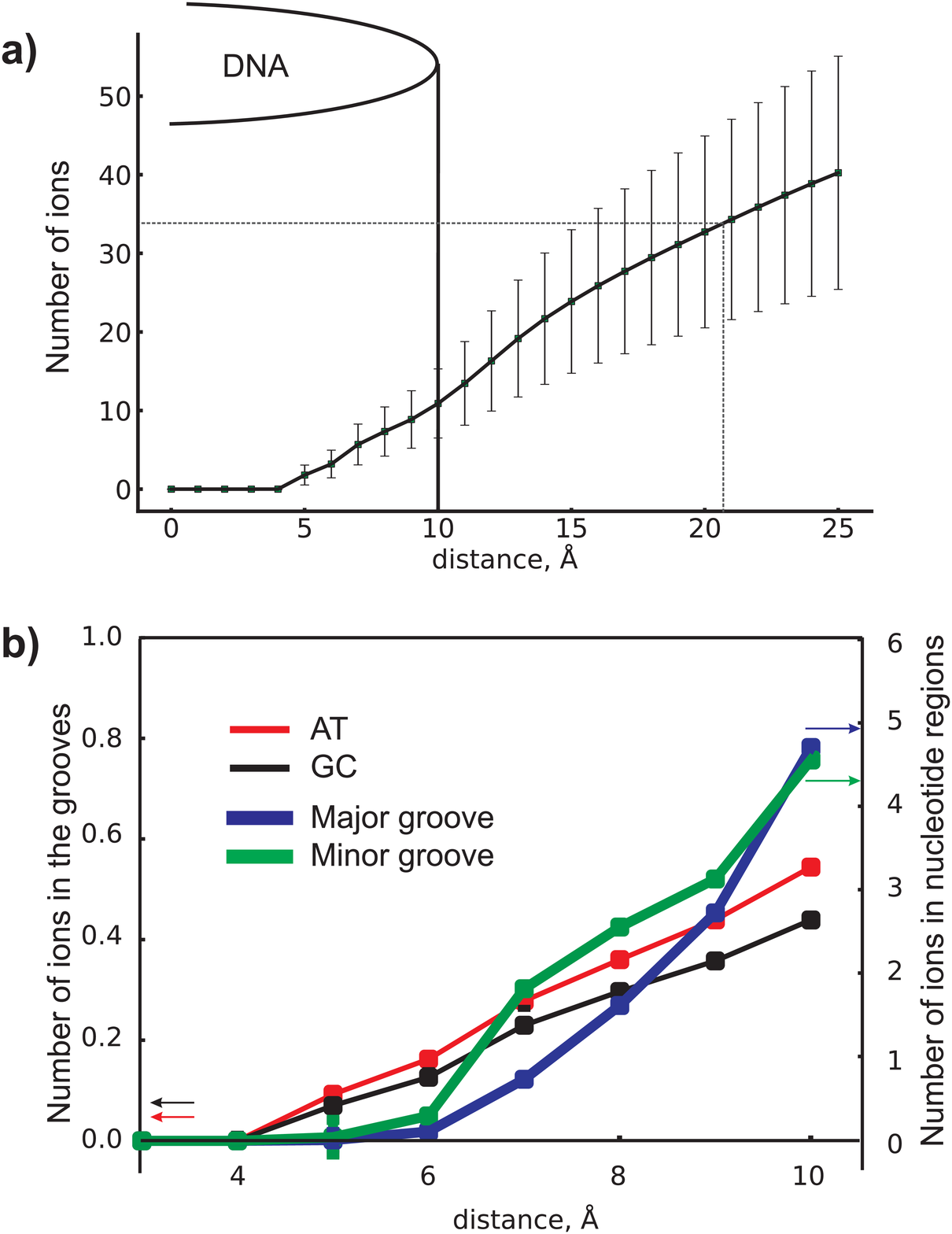}
}
\end{center}
\caption{Distribution of counterions around the DNA double helix. a) Distribution of counterions from the central axis of DNA macromolecule. The cylinder schematically shows the border of the DNA double helix. The dotted lines indicate the distance where the number of counterions corresponds to the factor 0.76 per one phosphate group. b) Distribution of counterions around the central axis of DNA macromolecule in the minor and major grooves of the double helix and in the nucleotide pairs.}
\end{figure}

To study the distribution of counterions around DNA, the number of K$^{+}$ ions was calculated for different distances from the helical axis (Figure 4a).  In our analysis, the center of the double helix was associated with the N$_1$ atom of purine nucleotide base (Guanine or Adenine). The results show that the number of counterions increases with the distance and reaches the value 0.76 per one phosphate group, predicted by Manning theory at the distance about 21 {\AA} from the center of the double helix (Figure 4a). This value is larger than those follows from formula (5). At the same time, the obtained results are within the distance interval provided by the model developed in the present work (Table 1). Note, the errors of the number of counterions, related to the fluctuation of the ions, are rather large and may be also estimated within the framework of Manning model \cite{manning_persistence_2006,manning_contribution_2006}.

In average, about 10 counterions are always present inside the DNA double helix in the minor and major grooves. The counterions in the minor groove are closer to the center of the double helix than in the case of the major groove, but in average the number of counterions within the phosphate radius of the DNA double helix is the same in the both grooves (Figure 4b). Thus, the  number of counterions inside the double helix is about 0.22 per one phosphate group that is in a good agreement with the results obtained from the model developed in the present work (Table 1). The number of counterions inside DNA obtained from the simulation is between $\gamma_{min}$ and $\gamma_{max}$ showing that the distribution of counterions in a cloud around the macromolecule is not uniform, and essential part of counterions is trapped inside the double helix.

To analyze the sequence effects, the number of counterions in A-T and G-C nucleotide pairs was calculated. The number of ions was analyzed for the slices perpendicular to the axis of the double helix that were determined as the regions between phosphorus atoms of nucleotides (Fig. 4b). The results show, that in average, the number of counterions in A-T nucleotide pairs is systematically higher than in the case of G-C nucleotide pairs. This conclusion may be explained by the natural narrowing of the minor groove in the A-T-rich nucleotide region making it more energetically favorable for the ion localization. Also, it matches the results of the previous molecular dynamics simulations of DNA with counterions \cite{Mocci,mocci2012,Lavery2014,Maddocks,Perepelytsya2018}.

The molecular dynamics simulations for DNA with K$^{+}$ counterions performed by other authors \cite{Maddocks} showed that the  distance where the negative charge of the DNA double helix is completely neutralized is located about  12   {\AA} from the macromolecule surface. Such a distance is close to the thickness of a counterion cloud around \emph{B}-DNA, estimated within the framework of the developed model (Table 1). According to the results of the work \cite{Maddocks}, the number of K$^+$ ions inside the double helix is about 0.2 per one phosphate group and essentially depends on a nucleotide sequence. This is close to the results of our MD simulations and model estimates (Table 1).

Thus, the obtained MD simulation results show that the distribution of counterions in a cloud around the DNA double helix is not uniform. The density of counterions inside the double helix is in general larger than in the region outside the macromolecule. Using the developed model on the basis of Manning counterion condensation theory, such effect may be explained as trapping of counterions inside the DNA grooves. The origin of such counterion trapping is the manifestation of specific interactions of K$^{+}$ with the DNA double helix.

\section{Conclusions}
The DNA macromolecule is a strong polyelectrolyte that condenses counterions from the solution. Condensed counterions may be localized in a cloud around the macromolecule and inside the internal compartments of the double helix (minor and major grooves). Taking into account the difference between counterions in the inner and outer regions of DNA, an analytical model for the description of trapping of monovalent counterions inside the grooves of the double helix was elaborated. The model of counterion condensation theory \cite{manning_1978} has been used as the basis. As a result, the range of concentrations of counterions that may be trapped inside the grooves of the double helix has been determined for the case of \emph{B}-DNA neutralized by monovalent ions. The estimated number of trapped counterions is within the range from 0.16 to 0.43 per one phosphate group. The number of counterions that may be localized in the internal compartments of the DNA double helix is much lower than the total number of condensed counterions. For the validation of the developed model the molecular dynamics simulation of \emph{B}-DNA with K$^{+}$ counterions has been carried out. The simulation data show that in average the number of the ions inside the grooves of the double helix is about 0.22 per one phosphate group that supports the results of the proposed model. Thus, the developed model describes general features of the structure of counterion cloud around DNA and is able to predict the number of counterions that are trapped inside the grooves of the double helix.\\
\\
S.Perepelytsya thanks Prof. Aatto Laaksonen and Prof. Francesca Mocci for discussion of the results. The present work was partially supported by the Project of the of the National Academy of Sciences of Ukraine (0119U102721).


\begin{thebibliography}{10}
\providecommand{\url}[1]{{#1}}
\providecommand{\urlprefix}{URL }
\expandafter\ifx\csname urlstyle\endcsname\relax
  \providecommand{\doi}[1]{DOI \discretionary{}{}{}#1}\else
  \providecommand{\doi}{DOI \discretionary{}{}{}\begingroup
  \urlstyle{rm}\Url}\fi

\bibitem{Saenger}
W.~Saenger, \emph{Principles of nucleic acid structure}. Springer-Verlag, New York, Berlin, Heidelberg, Tokyo (1984).

\bibitem{Blagoy}
Y.~Blagoy, V.~Galkin, G.~Gladchenko, S.~Kornilova, V.~Sorokin, A.~Shkorbatov,
  \emph{The Complexes of Nucleic Acids with Metal Cations in Solutions}.
  Naukova Dumka, Kyiv, (1991).

\bibitem{vologodskii_biophysics_2015}
A.~Vologodskii, \emph{Biophysics of {DNA}}. Cambridge University Press,
  Cambridge, (2015).
\newblock \doi{10.1017/CBO9781139542371}

\bibitem{Frank_Kamenetski__1987}
M.D. Frank-Kamenetski{\u{\i}}, V.V. Anshelevich, A.V. Lukashin, \emph{Soviet Physics
  Uspekhi} \textbf{30}(4), 317 (1987).
\newblock \doi{10.1070/PU1987v030n04ABEH002833}

\bibitem{levin_electrostatic_2002}
Y.~Levin, \emph{Reports on Progress in Physics} \textbf{65}(11), 1577 (2002).
\newblock \doi{10.1088/0034-4885/65/11/201}

\bibitem{kornyshev_structure_2007}
A.A. Kornyshev, D.J. Lee, S.~Leikin, A.~Wynveen, \emph{Reviews of Modern Physics}
  \textbf{79}(3), 943 (2007).
\newblock \doi{10.1103/RevModPhys.79.943}

\bibitem{manning_limiting_1969}
G.S. Manning, \emph{The Journal of Chemical Physics} \textbf{51}(3), 924 (1969).
\newblock \doi{10.1063/1.1672157}

\bibitem{oosawa_counterion_1970}
F.~Oosawa, \emph{Biopolymers} \textbf{9}(6), 677 (1970).
\newblock \doi{10.1002/bip.1970.360090606}

\bibitem{osawa_polyelectrolytes_1971}
F.~Oosawa, \emph{Polyelectrolytes}. M. Dekker, New York (1971).

\bibitem{manning_1978}
G.S. Manning, \emph{Quarterly Reviews of Biophysics} \textbf{11}(2), 179 (1978).
\newblock \doi{10.1017/S0033583500002031}

\bibitem{Das}
R.~Das, T.T. Mills, L.W. Kwok, G.S. Maskel, I.S. Millett, S.~Doniach, K.D.
  Finkelstein, D.~Herschlag, L.~Pollack, \emph{Phys. Rev. Lett.} \textbf{90}, 188103
  (2003).
\newblock \doi{10.1103/PhysRevLett.90.188103}

\bibitem{Andresen}
K.~Andresen, R.~Das, H.Y. Park, H.~Smith, L.W. Kwok, J.S. Lamb, E.J. Kirkland,
  D.~Herschlag, K.D. Finkelstein, L.~Pollack, \emph{Phys. Rev. Lett.}
  \doi{10.1103/PhysRevLett.93.248103}

\bibitem{tomic_dielectric_2007}
S.~Tomi\v{c}, S.D. Babi\v{c}, T.~Vuleti\v{c}, S.~Kr\v{c}a, D.~Ivankovi\v{c},
  L.~Gripari\v{c}, R.~Podgornik, \emph{Phys. Rev. E} \textbf{75}(2), 021905
  (2007).
\newblock \doi{10.1103/PhysRevE.75.021905}

\bibitem {perepelytsya_texture_2013}
S.~M.~Perepelytsya, G.~M.~Glibitskiy and S.~N.~Volkov,
\emph{Biopolymers} \textbf{99}, 508-516
(2013).
\newblock \doi{10.1002/bip.22209}

\bibitem{Liubysh2014}
O.~Liubysh, O.~Alekseev, S.Y. Tkachov, S.~Perepelytsya, \emph{Ukrainian Journal of
  Physics} \textbf{59}(5), 479 (2014).
\newblock \doi{10.15407/ujpe59.05.0479}

\bibitem{manning_persistence_2006}
G.S. Manning, \emph{Biophysical Journal} \textbf{91}(10), 3607 (2006).
\newblock \doi{10.1529/biophysj.106.089029}

\bibitem{manning_contribution_2006}
G.S. Manning, \emph{Biophysical Journal} \textbf{90}(9), 3208 (2006).
\newblock \doi{10.1529/biophysj.105.078865}

\bibitem{Katchalsky}
R.M. Fuoss, A.~Katchalsky, S.~Lifson, \emph{Proceedings of the National Academy of
  Sciences} \textbf{37}(9), 579 (1951).

\bibitem{Obukhov}
A. Deshkovski, S. Obukhov, M. Rubinstein, \emph{Phys. Rev. Lett.} \textbf{86}, 2341 (2001).
 \doi{10.1103/PhysRevLett.86.2341}

\bibitem{oshaughnessy_manning-oosawa_2005}
B.~O'Shaughnessy, Q.~Yang, \emph{Phys. Rev. Lett.} \textbf{94}(4), 048302
  (2005).
\newblock \doi{10.1103/PhysRevLett.94.048302}

\bibitem{trizac_onsager-manning-oosawa_2006}
E.~Trizac, G.~Téllez, \emph{Phys. Rev. Lett.} \textbf{96}(3), 038302 (2006).
\newblock \doi{10.1103/PhysRevLett.96.038302}

\bibitem{Mocci}
F.~Mocci, G.~Saba, \emph{Biopolymers} \textbf{68}(4), 471 (2003).
\newblock \doi{10.1002/bip.10334}

\bibitem{mocci2012}
F.~Mocci, A.~Laaksonen, \emph{Soft Matter} \textbf{8}(36), 9268 (2012).
\newblock \doi{10.1039/C2SM25690H}

\bibitem{Aksimentiev2012}
J.~Yoo, A.~Aksimentiev, \emph{The Journal of Physical Chemistry Letters}
  \textbf{3}(1), 45 (2012).
\newblock \doi{10.1021/jz201501a}

\bibitem{Lavery2014}
R.~Lavery, J.H. Maddocks, M.~Pasi, K.~Zakrzewska, \emph{Nucleic Acids Research}
  \textbf{42}(12), 8138 (2014).
\newblock \doi{10.1093/nar/gku504}

\bibitem{Canadian}
A.~Atzori, S.~Liggi, A.~Laaksonen, M.~Porcu, A.P. Lyubartsev, G.~Saba,
  F.~Mocci, \emph{Canadian Journal of Chemistry} \textbf{94}(12), 1181 (2016).
\newblock \doi{10.1139/cjc-2016-0296}

\bibitem{dans2016}
P.D. Dans, L.~Danilāne, I.~Ivani, T.~Dršata, F.~Lankaš, A.~Hospital,
  J.~Walther, R.I. Pujagut, F.~Battistini, J.L. Gelpí, R.~Lavery, M.~Orozco,
  \emph{Nucleic Acids Research} \textbf{44}(9), 4052 (2016).
\newblock \doi{10.1093/nar/gkw264}

\bibitem{Maddocks}
M.~Pasi, J.H. Maddocks, R.~Lavery, \emph{Nucleic Acids Research} \textbf{43}(4), 2412
  (2015).
\newblock \doi{10.1093/nar/gkv080}

\bibitem{Perepelytsya2018}
S.~Perepelytsya, \emph{Journal of Molecular Modeling} \textbf{24}, 171 (2018).
\newblock \doi{10.1007/s00894-018-3704-x}

\bibitem{perepelytsya2019positively}
S.~Perepelytsya, \emph{Ukrainian Journal of Physics} \textbf{65}(6), 500 (2020).
\newblock \doi{10.15407/ujpe65.6.510}

\bibitem{PV_EPJE_2007}
S.M. Perepelytsya, S.N. Volkov, \emph{The European Physical Journal E}
  \textbf{24}(3) (2007).
\newblock \doi{10.1140/epje/i2007-10236-x}

\bibitem{PV_EPJE_2010}
S.M. Perepelytsya, S.N. Volkov, \emph{The European Physical Journal E}
  \textbf{31}(2), 201 (2010).
\newblock \doi{10.1140/epje/i2010-10566-6}

\bibitem{PV_JML_2011}
S.~Perepelytsya, S.~Volkov, \emph{Journal of Molecular Liquids} \textbf{164}(1-2), 113 (2011).
\newblock \doi{10.1016/j.molliq.2011.04.015}

\bibitem{Bulavin}
L.A. Bulavin, S.N. Volkov, S.Y. Kutovy, S.M. Perepelytsya, \emph{Reports of the
  National Academy of Sciences of Ukraine} (10), 69 (2007).
\newblock ArXiv: 0805.0696

\bibitem{perepelytsya_BZ_2013}
S.~Perepelytsya, S.~Volkov, \emph{Journal of Physics: Conference Series}
  \textbf{438}(1), 012013 (2013).
  \newblock \doi{10.1088/1742-6596/438/1/012013}

\bibitem{perepelytsya_left_2013}
S.~Perepelytsya, S.~Volkov, \emph{Ukrainian Journal of Physics} \textbf{58}(6), 554
  (2013).
\newblock \doi{10.15407/ujpe58.06.0554}

\bibitem{nadassy_standard_2001}
K.~Nadassy, \emph{Nucleic Acids Research} \textbf{29}(16), 3362 (2001).
\newblock \doi{10.1093/nar/29.16.3362}

\bibitem{SPCE}
H.~Berendsen, J.~Grigera, T.~Straatsma, \emph{Journal of Physical Chemistry}
  \textbf{91}(41), 6269 (1987).
\newblock \doi{10.1021/j100308a0389}

\bibitem{1bna}
H.R. Drew, R.M. Wing, T.~Takano, C.~Broka, S.~Tanaka, K.~Itakura, R.E.
  Dickerson, \emph{Proceedings of the National Academy of Sciences} \textbf{78}(4),
  2179 (1981).
  \newblock \doi{10.1073/pnas.78.4.2179}

\bibitem{Charge2000}
K.J. McConnell, D.~Beveridge, \emph{Journal of Molecular Biology} \textbf{304}(5), 803
   (2000).
\newblock \doi{10.1006/jmbi.2000.4167}

\bibitem{rueda2004}
M.~Rueda, E.~Cubero, C.A. Laughton, M.~Orozco, \emph{Biophysical Journal}
  \textbf{87}(2), 800 (2004).
\newblock \doi{10.1529/biophysj.104.040451}

\bibitem{Ponomarev2004}
S.Y. Ponomarev, K.M. Thayer, D.L. Beveridge, \emph{Proceedings of the National
  Academy of Sciences} \textbf{101}(41), 14771 (2004).
\newblock \doi{10.1073/pnas.0406435101}

\bibitem{SJAF}
S.~Perepelytsya, J.~Uli\v{c}n\'{y}, A.~Laaksonen, F.~Mocci, \emph{Nucleic Acids
  Research} \textbf{47}(12), 6084 (2019).
\newblock \doi{10.1093/nar/gkz434}

\bibitem{Phillips}
J.C. Phillips, R.~Braun, W.~Wang, J.~Gumbart, E.~Tajkhorshid, E.~Villa,
  C.~Chipot, R.D. Skeel, L.~Kalé, K.~Schulten, \emph{Journal of Computational
  Chemistry} \textbf{26}(16), 1781 (2005)

\bibitem{MacKerell2000}
A.D. MacKerell, N.K. Banavali, \emph{Journal of Computational Chemistry}
  \textbf{21}(2), 105 (2000).
\newblock \doi{10.1002/(SICI)1096-987X(20000130)21:2<105::AID-JCC3>3.0.CO;2-P}

\bibitem{Foloppe2000}
N.~Foloppe, A.D. MacKerell, \emph{Journal of Computational Chemistry} \textbf{21}(2),
  86 (2000).
\newblock \doi{10.1002/(SICI)1096-987X(20000130)21:2<86::AID-JCC2>3.0.CO;2-G}

\bibitem{Humphrey}
W.~Humphrey, A.~Dalke, K.~Schulten, \emph{Journal of Molecular Graphics}
  \textbf{14}(1), 33 (1996)

\bibitem{SHAKE}
J.P. Ryckaert, G.~Ciccotti, H.J. Berendsen, \emph{Journal of Computational Physics}
  \textbf{23}(3), 327 (1977).
\newblock \doi{10.1016/0021-9991(77)90098-5}

\bibitem{TIP3P}
W.L. Jorgensen, J.~Chandrasekhar, J.D. Madura, R.W. Impey, M.L. Klein, \emph{The
  Journal of Chemical Physics} \textbf{79}(2), 926 (1983).
\newblock \doi{10.1063/1.445869}

\bibitem{Beglov}
D.~Beglov, B.~Roux, \emph{The Journal of Chemical Physics} \textbf{100}(12), 9050
  (1994).
\newblock \doi{10.1063/1.466711}

\bibitem{PME}
T.~Darden, D.~York, L.~Pedersen, \emph{The Journal of Chemical Physics}
  \textbf{98}(12), 10089 (1993).
\newblock \doi{10.1063/1.464397}

\bibitem{Zdorevskyi2020}
O.O. Zdorevskyi, S.M. Perepelytsya, \emph{The European Physical Journal E}
  \textbf{43}(12), 77 (2020).
\newblock \doi{10.1140/epje/i2020-12000-0}

\end{thebibliography}

\end{document}